\documentclass{aa-arxiv}

\usepackage{natbib}
\usepackage{graphicx}
\usepackage{siunitx}
\usepackage{txfonts}
\usepackage{lipsum}
\usepackage{subcaption}         
\usepackage{lscape}             
\usepackage{placeins}           
                                

\begin{document}

\title{Broadband heterodyne interferometry with a chirped femtosecond laser in the H-band}

\subtitle{}

   \author{F. Gudin\inst{1}\thanks{Corresponding author: felix.gudin@univ-cotedazur.fr}
        \and N. Forget\inst{1}
        }

   \institute{1. Université côte d’azur, CNRS, Institut de Physique de Nice (INPHYNI), France}
 
  \abstract
   {Infrared heterodyne interferometry offers a scalable alternative to direct interferometry for long-baseline telescope arrays. However, at near-infrared wavelengths, its sensitivity is limited by the electronic detection bandwidth and shot noise from the optical reference. Parallel detection via spectral multiplexing has long been identified as a potential means to increase the signal-to-noise ratio of heterodyne interferometers. We propose a novel heterodyne detection architecture based on highly dispersed broadband pulses from a mode-locked laser, fast InGaAs photoreceivers, and numerical correlation. This enables straightforward spectral multiplexing with commercial components to increase the signal-to-noise ratio (SNR), while extending instantaneous wavelength coverage to simultaneous J- and H-band operation. The scheme also supports high-resolution spectroscopic imaging ($R \simeq 10^{3}-10^{4}$). We derive the SNR of a two-arm heterodyne interferometer based on balanced photodetection and apply the model to the proposed scheme, using measurements from a single-spectral-channel, all-fiber, polarization-maintaining interferometer operating at 1.56~\si{\micro\meter} over an 8~\si{\nano\meter} bandwidth. We experimentally demonstrate an SNR exceeding 2 for a spectral flux density of \SI{188}{\pico\watt\per\nano\meter} at \SI{1.56}{\micro\meter}, with an integration time of \SI{0.4}{\milli\second}. Scaling the integration time and number of spectral channels suggests that fringe visibility of the brightest H-band stars could be achieved with \SI{1}{\meter\squared} telescopes within the typical atmospheric coherence time. These results represent a significant step toward broadband, scalable heterodyne interferometers, demonstrating the potential of ultrafast laser and telecommunications technologies for astronomical interferometry in the J+H bands. By combining broadband spectral multiplexing with numerical correlation, the architecture also opens a route to direct spectro-imaging without an additional spectrometer.
}


   \maketitle

\nolinenumbers
\section{Introduction}
\label{sec:introduction}

The quest to resolve ever-finer astrophysical structures, such as stellar surfaces, protoplanetary disks, and the immediate vicinities of black holes, has continually pushed the boundaries of angular resolution in astronomy. In the radio and submillimeter regimes, this pursuit has led to the development of Very Long Baseline Interferometry (VLBI, charbonneau2020), epitomized by the Earth-sized Event Horizon Telescope \citep{collaboration2019}, which has achieved resolutions of $\simeq$20\,µas at an observing wavelength of 1.3\,mm. By contrast, achieving comparable angular resolution, $(u,v)$-coverage or sensitivity at visible or infrared wavelengths over kilometer-scale baselines remains a formidable technical challenge.

To date, three primary detection schemes have been demonstrated for optical interferometry: direct detection of recombined beams (e.g., VLTI, CHARA \citep{tenbrummelaar2005, gies2024}, correlated heterodyne detection \citep{townes1984, hale2000, townes2008, michael2018, bourdarot2021}, and intensity correlation \citep{dravins2016, guerin2025}. While a detailed comparative analysis of these techniques lies beyond the scope of this paper, it is noteworthy that direct detection currently offers much superior sensitivity, albeit with constraints on baseline scalability and the number of telescopes that can be effectively integrated, essentially because the necessity to split and delay free-space beams with nanometric precision over hectometric distances.

Improving the sensitivity of detection techniques based on correlations, and especially optical heterodyne interferometry  - which the closest to VLBI in terms of architecture, phase sensitivity and scalability - might provide a path to solving the sensitivity bottleneck. However, as stated since the birth of heterodyne optical interferometry, the signal-to-noise ratio (SNR) of heterodyne detection at optical wavelengths inherently suffers from the fundamental shot-noise of the laser used as the local oscillator \citep{hale2000}. This effect, which is negligible at the radio and millimeter wavelengths, becomes the dominant noise source at optical wavelengths \citep{johnson2000}. Second, in contrast with direct detection schemes, the optical detection bandwidth of heterodyne detectors is limited by the electrical bandwidth of the detection chain rather than the sensitivity range of the detector itself. As a result, the detection bandwidth of heterodyne detection is extremely limited 
compared to that of direct recombination, even when taking into account the effect of the narrowband spectral filters ($\simeq$1\,THz) required, in practice, to increase the coherence length in optical interferometers \citep{glindemann2011}. Therefore, the SNR of heterodyne detection in the near-infrared is typically degraded by a large factor with respect to direct detection, limiting the faintest observable astronomical targets.

As pointed in particular in \citep{johnson2000}, the SNR of a single heterodyne detector is not improved by spectral multiplexing, that is, by combining several detection channels on a single detector. However, $N$ independent heterodyne detectors operating in $N$ different spectral channels increases the global SNR by a factor $\sqrt{N}$ \citep{fink1976}, suggesting that sensitivity improvements of $>10$ are within reach by multiplexing hundreds of heterodyne detectors \citep{monnier2018}. Spectral multiplexing also allows to extend the spectral coverage of the detection, while retaining multi-GHz, spectral resolution, which is the hallmark of heterodyne detection. Yet, massive spectral multiplexing raises its own technical challenges, among which are: (a) excess loss associated with wavelength division multiplexing (WDM), (b) matching the spectral width of the WDM channels with the detection bandwidth, (c) the complexity and challenge of managing a set of single-frequency lasers matched to the WDM grid, and (d) the large amount of signals to correlate.

Recent advances in telecommunications technology, propelled by high-speed data links operating at 40–100 Gb/s, have led to the development of cost-effective, high-performance components. These include photoreceivers with high quantum efficiency and/or fast response times, modulators, amplifiers, and high-speed digitizers, primarily in the J+H band but also extending into the K band (up to 2.2 µm). Such components could, at a reasonable cost and with manageable complexity, facilitate the creation of innovative architectures for parallel heterodyne signal detection, control, and correlation. Given that most of these devices rely on integrated or fiber-coupled optics, their performance depends critically on the efficient coupling of optical beams into single-mode polarization-maintaining optical fibers \citep{jovanovic2017}, an assumption underlying this work.

Regarding WDM losses, typical values are 0.5–1.5\,dB per channel for coarse channel spacings ($\simeq$10\,THz, CWDM) and 1.5–3\,dB for dense spacing (DWDM, 25–100\,GHz). These losses allow for up to 18 CWDM channels or 72 DWDM channels within the standard bands defined by ITU G.694.1/2. Despite the SNR penalty introduced by such losses, multiplexing >2 CWDM bands (or >4 DWDM bands) already improves the overall SNR. Additionally, the convergence of channel widths and electrical bandwidths of photoreceivers and modulators toward a few tens of GHz presents another advantage for heterodyne detection, assuming a properly spaced set of single-frequency lasers can be aligned with the ITU grids.
Optical frequency combs (OFCs) with large line spacing (10-100 GHz \citep{torres-company2014} hold the promise to exhibit such properties at the cost of having to reference the comb to align with the WDM channels. In the context of heterodyne detection, OFCs, which produce train of evenly delayed and co-phased pulses, also provide means to stabilize optical path lengths but also time tag and synchronize data acquisition systems. However, OFSc tied to the ITU standards and covering the H bands are not yet widely available and the search for compact, cost-effective and robust astrocombs remains an active field of research.

In this work, we describe an alternative WDM scheme which does not require a referenced optical frequency comb for multi-channel heterodyne detection in the H-band. In this scheme, coined as \emph{ultrafast serrodyne} detection, the local oscillator is a train of broadband laser pulses exhibiting a instantaneous frequency changing linearly over time. The name serrodyne combines the Latin \emph{serra} ("saw"), referring to sawtooth modulation \citep{johnson2010}, with the suffix \emph{-dyne}, by analogy with heterodyne, to denote a frequency-shifting technique based on deterministic phase modulation. We show that, for an adequate choice of periodicity and frequency drift rate, such a laser source may drive simultaneously several spectral WDM channels, without the need for spectral referencing. As a proof of concept, we experimentally demonstrate two-arm, single-channel, 44-GHz serrodyne interferometer and assess the SNR of such an interferometer with a incoherent source at 1560\,nm. 

The paper is organized as follows. We first introduce the principles of serrodyne detection. We then derive the expected properties of heterodyne and serrodyne correlation measurements in the context of balanced detection and present, as a proof of concept, a fully fibered two-arm interferometer. Using an attenuated superluminescent diode as the light source, we investigate the correlation contrast as a function of the input power, optical path mismatch, and integration time. Finally, we discuss the extension of the technique to multiple CWDM channels, its scalability to a larger number of interferometric arms, and its potential applications to broadband spectro-imaging.

\section{Broadband heterodyne interferometry}
\label{sec:boadband_heterodyne}

\subsection{Heterodyne interferometer principle}

In optical heterodyne interferometry, the optical field of a distant unresolved  astronomical object is collected by separated telescopes and mixed with a strong reference laser field, called local oscillator (LO). The beating between the fields is recorded by a fast photodetector at each telescope. Under standard assumptions of linear detection and LO stability, the time-averaged correlation of the two beatings is proportional to the mutual coherence
function of the signal fields at the two telescopes, and hence to the complex visibility $V(b,\nu)$ at the projected baseline. 
\begin{figure}[ht!]
   \centering
   \includegraphics[width=\hsize]{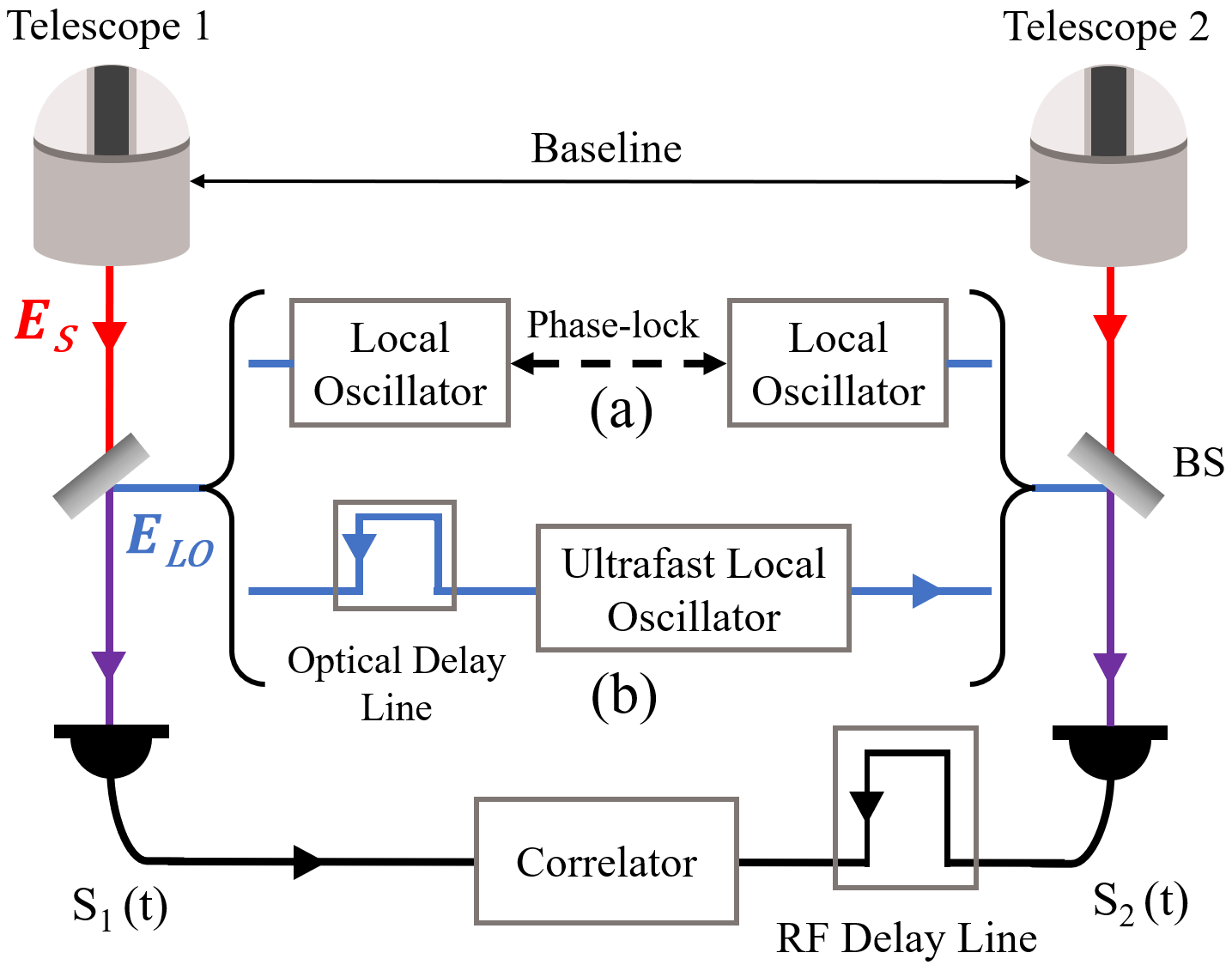}
      \caption{ (a) Conventional and (b) the proposed optical heterodyne interferometry setup. BS: beam splitter}
         \label{fig:heterodyne}
\end{figure}
Phase control of the LO at each station, illustrated in Fig.~\ref{fig:heterodyne}(a), provides an absolute phase reference, allowing recovery of both the amplitude and phase of $V(b,\nu)$.
For a spatially incoherent source, the van Cittert--Zernike theorem states that the complex visibility is the Fourier transform of the intensity distribution of the astronomical source. Thus, measuring $V(b,\nu)$ over many baselines and, if desired, multiple spectral channels, provides discrete samples of the two-dimensional Fourier transform of the stellar brightness distribution. Heterodyne detection enables measurements via standard RF techniques rather than direct optical-fringe detection. Upon electrical filtering of the DC contributions, only the beat signal remains, whose amplitude and phase are related to the optical field to within a known and controlled LO amplitude and phase. A cross-correlation is then computed between the RF beat signals recorded at two telescopes. 
In this context, heterodyne detection circumvents the challenges of long-distance weak-light
transport and the stringent optical-path stabilization required by direct interferometry. 
\subsection{Serrodyne local oscillator}
\label{Serrodyne principle}
\begin{figure}[ht!]
   \centering
   \includegraphics[width=\hsize]{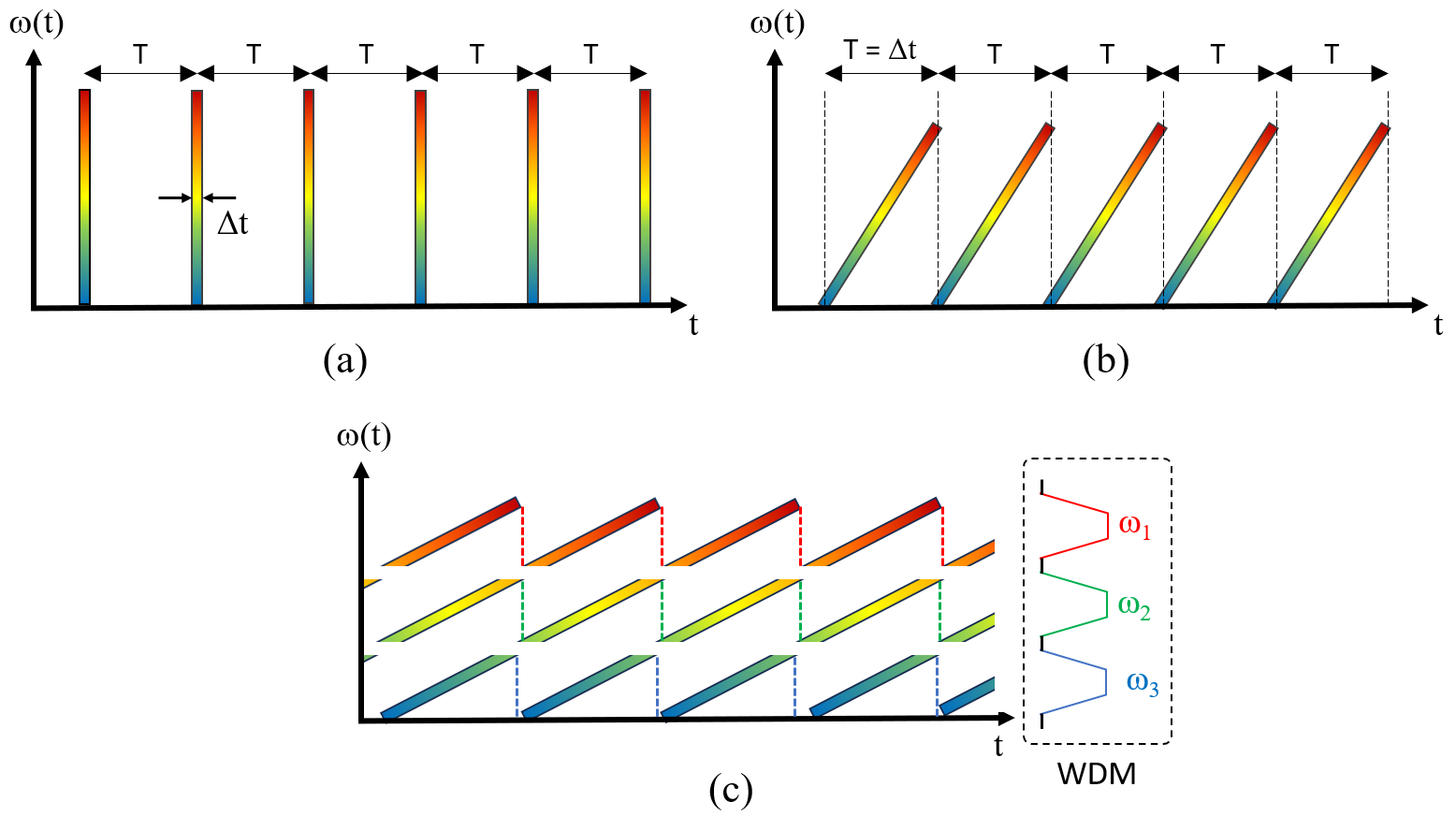}
      \caption{Time--frequency diagrams illustrating the spectral content of the LO. (a) Train of ultrashort pulses emitted by a mode-locked femtosecond laser, for which all spectral components within each pulse arrive simultaneously. (b) Serrodyne pulse train obtained after dispersion, acting as a swept source with a linear drift of the instantaneous frequency. (c) Recovery of three independent serrodyne channels by wavelength-division multiplexing (WDM) of the overlapping swept-source signals.}
    \label{fig:Freq-time representation}
\end{figure}

In contrast to conventional optical heterodyne interferometry, the alternative scheme in Fig.~\ref{fig:heterodyne}(b) employs a LO that is not a continuous-wave single-frequency laser but a periodic train of pulses exhibiting a linear drift of the instantaneous frequency.
In practice, such an LO is derived from a mode-locked femtosecond laser emitting a train of ultrashort pulses, represented in Fig.~\ref{fig:Freq-time representation}(a), where $f_{\mathrm{rep}}$ is the repetition rate, $T=1/f_{\mathrm{rep}}$ the period of the pulse train, $\Delta\omega$ the optical bandwidth of the LO, and $\Delta t$ the pulse duration. Applying chromatic dispersion to temporally stretch the pulses up to the repetition period ($\Delta t = T$) results in a regime where the instantaneous frequency $\omega(t)$ follows a sawtooth temporal evolution, as illustrated in Fig.~\ref{fig:Freq-time representation}(b). As demonstrated in laser metrology \citep{deschenes_chirped_2015}, this so-called chirped pulse heterodyne can be noise-equivalent to a CW local oscillator heterodyne experiment, where the stretched pulse train effectively supplies its full spectrally available power.
When the swept source undergoes additional dispersion such that $\Delta t > T$, successive pulses overlap temporally and optical frequencies originating from different pulses become simultaneously present. Therefore, in Fig.~\ref{fig:Freq-time representation}(c), wavelength-division multiplexing can be used to recover independent serrodyne channels from the overlapping signals by partitioning the broadband LO into adjacent spectral bands, thereby enabling parallel serrodyne detection.

A strength of the serrodyne signals is that there is no alignment requirement (spacing and offset) with regard to WDM grids because of the continuous frequency drift. Another advantage is that, providing that the $\Delta \omega/\Delta t$ factor may be adjusted, which can be easily done as described in Section~\ref{sec:setup},   different WDM spacings can be addressed with the same LO. Last, the signal is periodic with a period $T$ on all channels, which means that delay lines in Fig.\ref{fig:heterodyne}(b) of optical path lengths of a most $c \times T$ (typically a few meters) are required to synchronize the LO at different telescopes. Additionally, pulse shapers \citep{weiner2011} offer a convenient and integrated platform for precise control over both the phase and amplitude of the local oscillator (LO) pulses.

\section{Balanced heterodyne detection}

\subsection{General formulation}
\label{sec:heterodyne_theory}
The incoming stellar electric field $\mathcal{E}_s(t)$
interferes with a laser field  $\mathcal{E}_{\mathrm{LO}}(t)$ on a fast photodiode characterized by a impulse response $B(t)$, producing a photocurrent proportional to the incident intensity. 

\begin{figure}[ht!]
   \centering
   \includegraphics[width=\hsize]{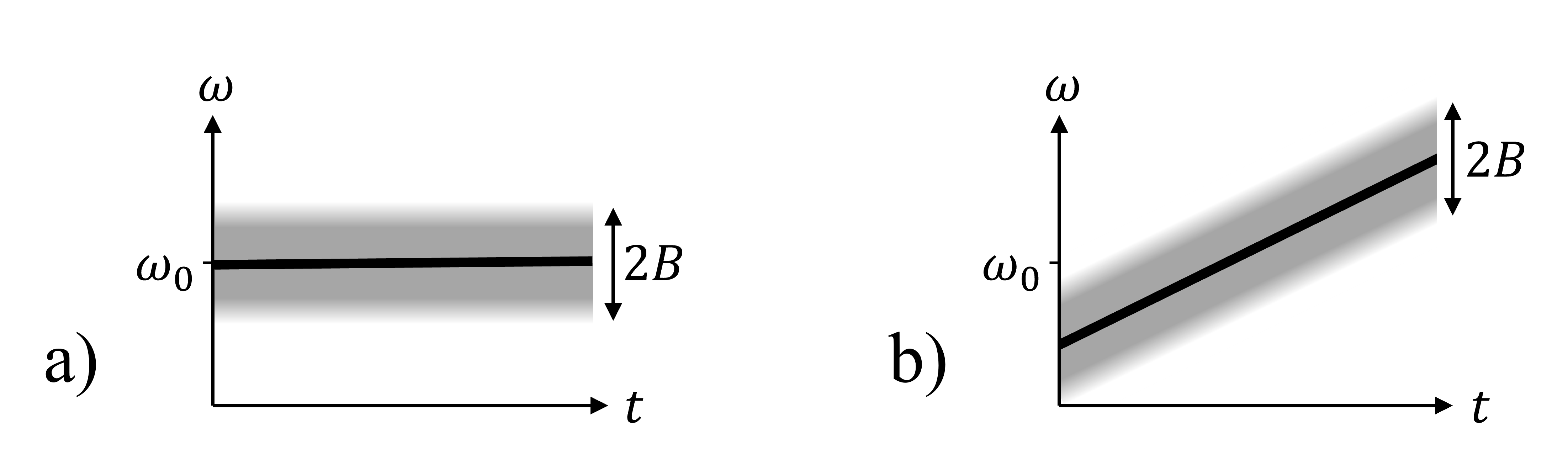}
      \caption{Comparison between (a) conventional continuous-wave single-frequency and (b) serrodyne heterodyne detection. Black: schematic time--frequency representation of the local oscillator (LO). Grey: optical bandwidth covered by the heterodyne detection, where $B$ denotes the photodetector bandwidth.}
         \label{fig:principle}
\end{figure}
Let $\omega_0$ be the common angular frequency of both the signal and LO fields, such that $\mathcal{E}_s(t)=E_s(t)e^{i\omega_0 t}$ and $\mathcal{E}_\mathrm{LO}(t)=E_\mathrm{LO}(t)e^{i\omega_0 t}$. Let $P_k(t) = |E_k(t)|^2$ be the average optical power incident on the photodetector 
for, respectively, the local oscillator ($k=\mathrm{LO}$) and the stellar source ($k=S$). Assuming a perfect 50/50 beam combiner, the total incident field on a photodetector located at one of the two outputs of the beam splitter is $\frac{1}{2}(\mathcal{E}_\text{LO}\pm\mathcal{E}_s)$ and the measured output voltage $S_\pm(t)$ writes:
\begin{equation}
\label{eq:S_def}
    S_\pm(t) = \eta\left( \frac{1}{2}\overline{P_{\mathrm{LO}}}(t) + \frac{1}{2}\overline{P_s}(t)
    \pm \text{Re}\{\overline{E_\mathrm{LO}E_s^*}\}(t) \right)
\end{equation}
where $\pm$ is related to the 50/50 output port and the upper bar stands for the convolution by the temporal impulse response $B(t)$ of the photodiode. All terms related to the photodetector's sensitivity, the transimpedance amplifier gain, optical or electrical losses are contained in the proportional factor $\eta$ (unit: V/W).  
The first and second terms as proportional to the optical intensities of the LO and stellar source as measured independently, the second being much weaker ($\ll$nW) than the first one ($\simeq$mW). The third term, also referred to as the cross-term of the intermediate frequency signal carries the instantaneous phase of the stellar field, with respect to the instantaneous phase of the LO, at the telescope focus. Because of the convolution by $B(t)$, no phase information beyond the detection bandwidth of the photodetection chain can be recorded nor retrieved. 

By using an ideal balanced detection between the two outputs of the 50/50, the differential signal is simply:
\begin{equation}
\label{eq:S2_def}
    S(t) = S_+(t)-S_{-}(t) = N(t) + 2\eta \text{Re}\{\overline{E_\mathrm{LO}E_s^*}\}(t)
\end{equation}
where $N(t)$ stands for an additive noise source, which is the sum of all non-common noise sources of the two photodiodes. At optical wavelengths, the shot noise of the LO is the dominant noise source. As all of the observables are linked to statistical values, the time average, over a time interval $T$, of a component $f$ will be noted $\langle f(t) \rangle$.
Since $E_S(t)$ describes the electric field of a thermal source within a given optical bandwidth $\Delta\nu \gg B$, its phase varies randomly with a characteristic time $\simeq$$1/\Delta\nu$ $\ll 1/B$, so that $\langle S(t) \rangle \simeq 0$.

By assuming a LO field of constant time amplitude, its expression may be written as $E_\mathrm{LO}(t)= \left|E_\mathrm{LO}^0\right| \exp(-i\phi(t))$. 
With two telescopes equipped with identical photodetectors and receiving the same stellar optical power, the output electric signals ($S_1$ and $S_2$) have the same expressions but the electric fields may be dephased and delayed because of different optical paths. Here, we index the field by the subscripts $1$ and $2$ to distinguish the fields at telescopes 1 and 2 respectively. 
\begin{equation}\label{eq:balanced2}
    \left\{
    \begin{aligned}
        S_1(t) &= N_1(t) + 2 \eta\left|E_\mathrm{LO}^0\right| \text{Re}\{\overline{E_{S,1}(t)e^{i\phi_1(t)}}\}
        \\S_2(t)  &= N_2(t) + 2\eta\left|E_\mathrm{LO}^0\right|\text{Re}\{\overline{E_{S,2}(t)e^{i\phi_2(t)}}\}
    \end{aligned}
    \right.
\end{equation}

\subsection{Pearson correlation coefficient}
In the context of numerical correlation, the degree of correlation between $S_1(t)$ and $S_2(t)$ can be assessed by computing the Pearson correlation coefficient over a time interval $T$, defined as:
\begin{equation}
\label{eq:rho_def}
\rho_{1,2}(\tau) =
\langle S_1(t)\,S_2(t-\tau)\rangle/
\sqrt{\langle S_1^2(t)\rangle\,\langle S_2^2(t-\tau)\rangle},
\end{equation}
The choice of a dimensionless metric, equivalent to the normalized cross-correlation in radioastronomy, is comparable across different baselines or observing conditions. When the additive noises can be neglected, the statistical independence of the fluctuations of the LO and SLED field, Eq.\ref{eq:rho_def} reduces to the normalized temporal autocorrelation of the signal field:
\begin{equation}
\label{eq:rho_field}
\rho_{1,2}(\tau)
\approx
\frac{\operatorname{Re}\!\left\langle \overline{E_{S,1}(t)}\,\overline{E_{S,2}^{*}(t-\tau)}\right\rangle}
{\left\langle |E_S(t)|^2 \right\rangle}.
\end{equation}
Equation~\eqref{eq:rho_field} shows that \(\rho_{1,2}(\tau)\) is the real part of the normalized first-order temporal coherence function \(g^{(1)}(\tau)\) of the optical sources, and that \(\rho_{1,2}(0)\) is the real part of the interferometric fringe visibility $\mathcal{V}$. As described in \citep{hale2000}, the full complex visibility can be retrieved by introducing an additional phase modulation and a lock-in detection.

In practice, the effect of the additive noise cannot be neglected and a more realistic model is required to understand the experimental observations. We assume that $N_1$ and $N_2$ are independent real-valued stationary random variables ($\langle N_1 N_2\rangle \simeq 0$) of variance $\sigma_N^2$. $\sigma_N^2$ corresponds to the average electrical power of the random voltage fluctuations in the absence of stellar light. 
We also assume that that $\overline{E_{S,1}}$ and $\overline{E_{S,2}}$ are complex-valued random variables, independent from $N_1$ and $N_2$, but mutually partially correlated. The variance of $\eta|E_\mathrm{LO}^0\overline{E_{S,1}}|$ and $\eta|E_\mathrm{LO}^0\overline{E_{S,2}}|$, noted $\sigma_S^2$, is common to the two telescopes. It represents the average electrical power of the beating between the LO and the stellar light. 
Under these assumptions, combining Eq.\ref{eq:balanced2} and \ref{eq:rho_def}, yields the following expression for the correlation coefficient:
\begin{equation}\label{eq:rho12}
    \rho_{1,2}(\tau) = \text{Re}\left\{\frac{\eta^2P_\mathrm{LO}^0\langle \overline{E_{S,1}^{}(t)e^{i\phi_1(t)}}\,\overline{E_{S,2}^*(t-\tau) e^{-i\phi_2(t-\tau)}} \rangle}{\sigma_S^2 + \sigma_N^2/2} \right\}
\end{equation}
The numerator of Eq.\ref{eq:rho12} is the real part of the cross-correlation function of the stellar fields (convoluted by $B(t)$) at telescopes 1 and 2, with phase offsets $\phi_1(t)$ and $\phi_2(t-\tau)$. The effects of the variation of the relative optical path between the stellar source and the telescopes as well as the atmospheric turbulence are contained in the phases of $E_{S,1}$ and $E_{S,2}$. whereas phase variations of the LO and of the link uses to distribute the LO are contained in $\phi_1$ and $\phi_2$. From Eq.\ref{eq:rho12}, it follows that all these phase fluctuations may be treated as a single local phase contribution.   

To link $\sigma_S$ and $\sigma_N$ to the input optical powers, some additional assumptions are required. First, we restrict the noise source to the shot noise of the LO, which is relevant for balanced detection. For an ideal laser at frequency $\nu_0$, the optical shot noise gives rise to random fluctuations of $\overline{P_\mathrm{LO}}(t)$, which are statistically independent on all photodiodes, and characterized by $\sigma^2_N \simeq \eta^2 P_\mathrm{LO}\times h\nu_0 B$. This can be interpreted as the beat term between the classical field $\mathcal{E}_\mathrm{LO}$ and the zero-energy point quantum field \citep{loudon2000}, which is half a photon per unit of detection bandwidth. Second, we suppose that the stellar source may be treated as thermal light, characterized by a Poisson distribution for which the standard deviation of the power fluctuations equals the average optical power. At the output of the photodiode, this leads to $\sigma^2_S \simeq \eta^2 P_S P_\mathrm{LO}\times B/\Delta\nu$, where the factor $B/\Delta\nu$ accounts for the temporal averaging of the fast optical fluctuations (characteristic timescale of $\simeq1/\Delta\nu$) by the finite response time of the photodiode ($\simeq1/B$).
\subsection{Maximum correlation and background level}
\label{sec:max_visibility}
To assess the effect of the noise term in the denominator of Eq.\ref{eq:rho12} it is useful to consider the case of the zero base line ($E_{S,1}=E_{S,2}$) with constant $\phi_1$ and $\phi_2$ phase functions and $\tau = 0$, that is, of the maximum visibility. In this case, the expression of Eq.\ref{eq:rho12} simplifies in:
\begin{equation}\label{eq:rho12_zero}
\rho_{1,1}(0) \simeq \frac{\sigma_S^2}{\sigma_S^2 + \sigma_N^2/2} = \frac{P_S/\Delta\nu}{P_S/\Delta\nu + h\nu_0/2} 
\end{equation}
For a sufficiently large power spectral density $P_S/\Delta\nu$, the maximum visibility tends towards unity. For a power spectral density equal to half a LO photon energy  $\rho_{1,1}(0)$ drops to $1/2$. For weaker stellar power density, the maximum visibility tends towards $0$. 

The approximation made so far eventually breaks down when the statistical fluctuations of the $\langle N_1 N_2\rangle$ term become comparable to or larger than the stellar contribution to the numerator of Eq.~\ref{eq:rho12}. Beyond this point, the stellar signal cannot be reliably distinguished from the background value of $\rho_{1,1}(0)$ measured without stellar light. Since $\langle N_1 N_2\rangle\simeq \sigma_N^2/\sqrt{B\,T}$, this background value is $2/\sqrt{B\,T}$, whatever the physical noise sources. 
For a shot-noise-limited measurement the stellar flux for which the signal-to-noise of the correlation equals $1$ corresponds to $P_{S,min}/\Delta\nu = h\nu_0/\sqrt{B\,T}$. This result, which holds for a continuous-wave single-frequency LO, is identical to that of \citep{hale2000} for example.

\section{Serrodyne detection}\label{sec:serrodyne}
In serrodyne detection, $\phi_1(t)$ and $\phi_2(t)$ are not constant functions but describe the frequency drift of the local oscillator. For a purely linear frequency drift, the instantaneous phase drift, with respect to the carrier $\nu_0$ is $\phi_j(t) =2\pi\left(\nu(t)-\nu_0\right)\,(t-\tau_j) =  t (t -\tau_j)/\varphi_2$ where $\varphi_2$, the chirp coefficient, quantifies the speed of the frequency drift and $\tau_j$ is the time delay at the origin of time at telescope $j=1,2$. For the sake of clarity, let $\tau_1 = \tau_2 = 0$, which can be achieved by adjusting the group delay of the LO at both telescopes.

Because the phase drift is slow compared with the characteristic timescale of the photodetection chain ($1/B$), the phase factors containing $\phi_1(t)$ and $\phi_2(t)$ in Eq.~\ref{eq:rho12} vary negligibly during the detection interval. They can thus be taken outside the time average (ie out of the upper bar operator). The phase difference term $\phi_1(t)-\phi_2(t-\tau)$ is therefore equal to $2t\tau/\varphi_2 - \tau^2/\varphi_2$. However, this phase variation is too fast with respect to $T$ and cannot be factorized from the bracket ($\langle\rangle$ in Eq.~\ref{eq:rho12}).
A first conclusion is that this term cancels for $\tau = 0$: serrodyne detection detection is undistinguishable from heterodyne detection with a single-frequency LO when $\tau = 0$.

If $\tau\neq 0$, then the phase difference is a linear function of time and the time average at the numerator of Eq.\ref{eq:rho12} turns into a windowed Fourier transform of the mutual correlation:
\begin{equation}\label{eq:rho12_serrodyne}
    \rho_{1,2}(\tau) = \text{Re}\left\{\frac{\eta^2P_\mathrm{LO}^0\langle \overline{E_{S,1}^{}(t)}\,\overline{E_{S,2}^*(t-\tau)} e^{i 2 t \tau/\phi_2}\rangle }{\sigma_S^2 + \sigma_N^2/2} \,e^{-i \tau^2/\phi_2}\right\}
\end{equation}
The inverse Fourier transform of $\rho_{1,2}(\tau)$ is therefore the optical spectrum of the stellar source. This feature is quite distinctive from the case where the LO is a single-frequency laser. Serrodyne detection is akin to white-light interferometry where the LO spectrum plays the role of the bandpass filter. As for direct recombination, $\tau=0$ must be actively maintained to recover the fringe visibility at a given baseline. However, the required accuracy is of the order of a fraction of the coherence time (or coherence length) of the LO rather than the optical period of the LO. 

\section{Proof of concept - single channel correlator}
\label{sec:poc}

In this section, the feasibility of this new approach is demonstrate with an all-fiber, single-channel serrodyne detection of an
attenuated superluminescent diode (SLED) over a bandwidth of \SI{8}{\nano\meter} at
\SI{1.56}{\micro\meter}.
Section~\ref{sec:setup} describes
the setup of the heterodyne detection interferometer. In Section~\ref{sec:measurement}, we present
the temporal correlation measurement that yields the inverse Fourier transform of the temporal coherence function of the source. In Section~\ref{sec:detection_limit} and Section~\ref{sec:shot_noise}, we assess the sensitivity of
this single-channel detection system in relation to the shot-noise of the LO.

\subsection{Serrodyne experimental setup}
\label{sec:setup}

\begin{figure}[ht!]
\centering
\includegraphics[width=\hsize]{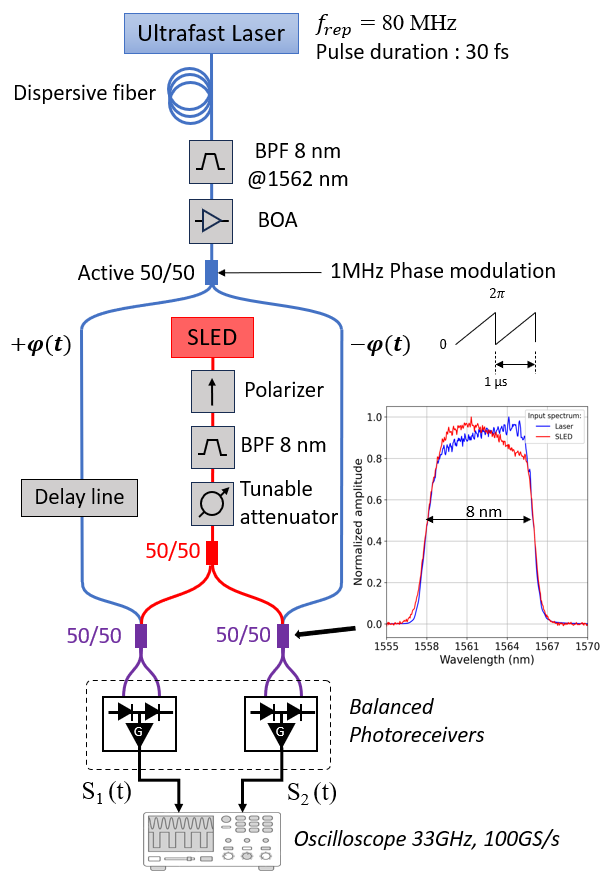}
\caption{All-polarization-maintaining fiber interferometer setup. $f_{rep}$: pulse repetition rate; Delay line: motorized free-space stage ; BPF: 8 nm bandpass filter centered at 1562\ nm; BOA: PM Booster Optical Amplifiers,  SLED: superluminescent diode ; Oscilloscope: Tektronix DPO70000SX. Normalized spectra of the SLED and LO at the balanced photoreceivers input.}
\label{fig:setup}
\end{figure}
Our prototype described in Fig.~\ref{fig:setup}, is a two-arm, single-spectral channel, polarization-maintaining heterodyne interferometer built from telecom fiber components in the H band. The serrodyne LO is derived from a commercial Toptica mode-locked Erbium-doped oscillator delivering a train of broadband ($\simeq$200\,nm) and short pulses ($\simeq$20\,fs) at a repetition rate of \SI{80}{\mega\hertz}. 
The pulse durations of the LO are temporally stretched to match the repetition period (\SI{\approx 12.5}{\nano\second}) by propagating through a dispersive single-mode fiber, providing a total accumulated chromatic dispersion of \SI{750}{\pico\second\per\nano\meter}.  The artificial stellar source consists of a continuous-wave, broadband (FWHM of \SI{60}{\nano\meter}) superluminescent diode (SLED), whose optical power is adjusted using a variable attenuator.
The spectral widths of the LO and SLED are matched using two tunable fiber-coupled band-pass filters (BPFs). For the fixed accumulated chromatic dispersion, \SI{8}{\nano\meter}-wide spectra (see Fig.~\ref{fig:setup}) prevent temporal overlap between consecutives pulses, thereby ensuring single-channel operation. 
The resulting quasi-continuous serrodyne LO is first amplified (BOA1004P, Thorlabs) and then split into two arms (arm 1 and arm 2) 
by a multifunction integrated optical chip (MIOC-1550-22-PG, Optilab). This device comprises a polarizer, a Y-junction coupler, and a dual electro-optic phase modulator. This phase modulator adds a linear phase shift of $2\pi F t$  between the two LO arms at $F=$\SI{1}{\mega\hertz}. In addition to simulating the effect of the Earth's rotation, it provides a means of separating the heterodyne signal from technical noise sources, as done in \citep{hale2000}. Arm 1 includes a home-built motorized free-space delay line providing a \SI{400}{\milli\meter} travel range with a minimum step size of \SI{50}{\nano\meter}. The fiber length of arm 2 is adjusted to match the reference optical path of arm 1. The LO is then mixed with the SLED by two broadband 50:50 fiber couplers and detected by two \SI{22}{\giga\hertz} amplified balanced photodetectors. Each detector incorporates two matched photodiodes and a differential amplifier that subtracts the photocurrents, thereby suppressing common-mode noise sources, such as laser relative intensity noise (RIN) and the DC background, while preserving the heterodyne beat signal. 

The optical path lengths from the 50:50 coupler to the photodiodes are matched to within a few micrometers, limiting the differential delay (or skew) to only a few tens of femtoseconds.
The resulting electrical signals pass through RF high-pass filters (1\,GHz) before digitization by a
\SI{33}{\giga\hertz} oscilloscope. By varying the optical path difference between the two arms, we scan the delay \(\tau\) and experimentally determine \(\rho_{1,2}(\tau)\). At each delay, the modulus of the Pearson coefficient is computed from the Fourier component at \SI{1}{\mega\hertz} of the Eq.\ref{eq:rho_def}. This pseudo-lock-in detection effectively suppresses low-frequency mechanical noise, including transient oscillations of the motorized delay line during closed-loop settling to the target position, as well as external mechanical vibrations propagating through the optical fibers.

\subsection{Correlation contrast as a function of the LO group delay}
\label{sec:measurement}

\begin{figure}[ht!]
\centering
\includegraphics[width=\hsize]{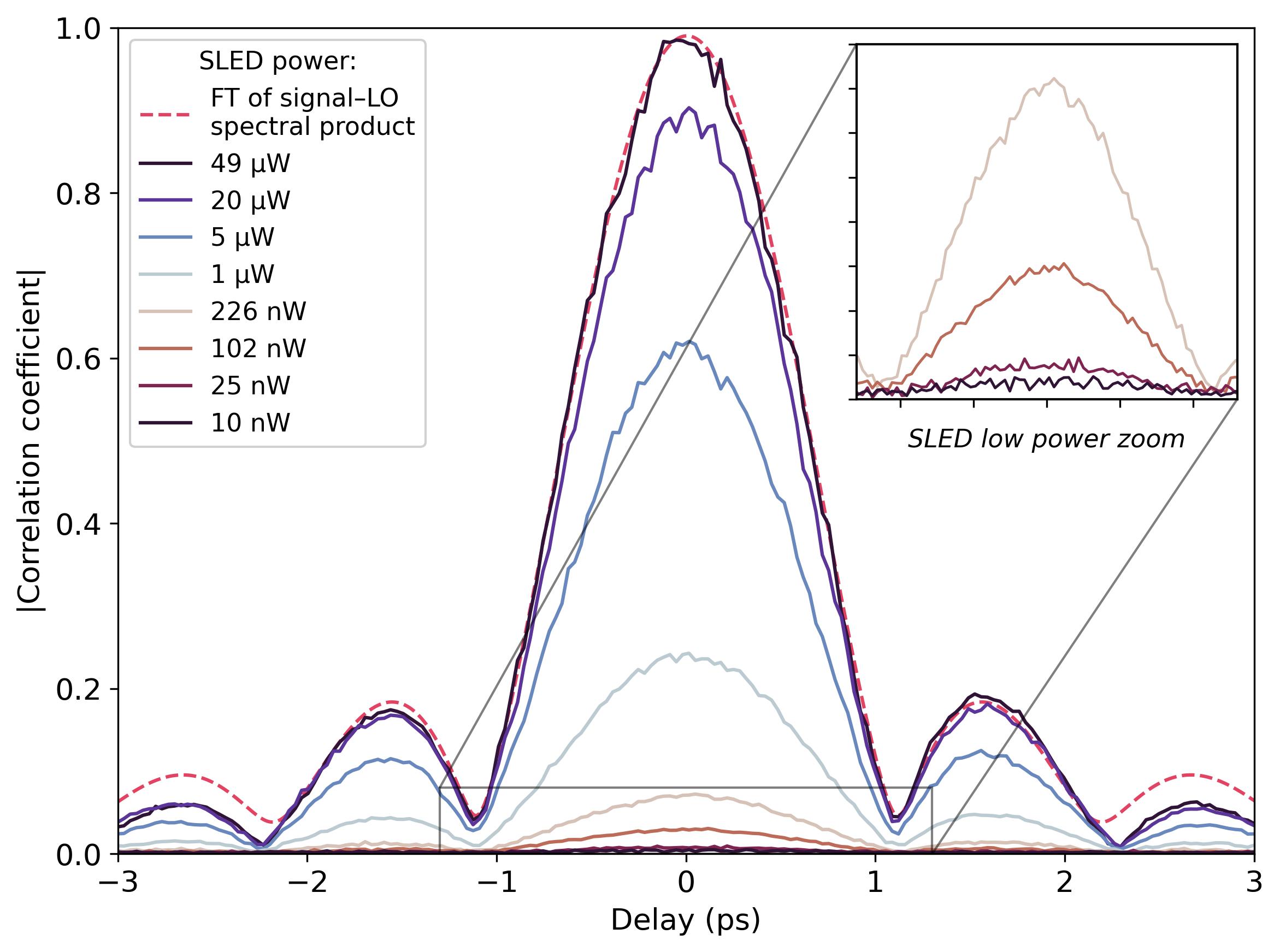}
\caption{Temporal correlation functions measured for different SLED power levels. At each delay, the Pearson correlation coefficient is computed from acquisitions of \SI{100}{\micro\second} duration sampled at \SI{100}{GS/s}. The SLED powers are measured after the tunable attenuator and can be directly compared to the total stellar flux coupled into a single-mode fiber at the telescope focus.}
\label{fig:corr_100us}
\end{figure}

To verify Equation~\ref{eq:rho12_serrodyne}, we first vary the optical path difference between the two LO arms and compute the Pearson correlation coefficient $\rho_{1,2}(\tau)$ at each delay $\tau$ for different SLED input powers. For practical reasons, the integration time of each measurement is set to \SI{100}{\micro\second}, which corresponds, for a sampling rate of 100\,GS/s, to $2\times10^7$ samples for the two channels. As described in Section \ref{sec:heterodyne_theory}, $|\rho_{1,2}(\tau)$|, that is the mutual coherence function of the LO and star fields, is equivalent to measuring the modulus of the Fourier transform of the spectral overlap between the LO and star light. Given the top-hat spectral distributions of both the SLED source and the LO shown in the inset of Fig.~\ref{fig:setup}, $|\rho_{1,2}(\tau)$| is expected to exhibit a $\mathrm{sinc}^2$-like profile (Eq~\ref{eq:rho12_serrodyne}). The expected background floor of $\rho_{1,2}$ being $2/\sqrt{B T}$, a value at $\simeq1.5\,$10$^{-3}$ is also expected.
Fig.~\ref{fig:corr_100us} displays the experimental $|\rho_{1,2}$| as a function of $\tau$ for different SLED power as well as the square modulus of the Fourier transform of $|E_\mathrm{LO}(\omega)E_s(\omega)|^{1/2}$ (red dashed curve). 

In the strong SLED regime, excellent agreement is observed between the measured signal and the calculated Fourier transform. Imperfections in the balanced detection system and/or the 50/50 beam splitters likely limit the fringe visibility to values below unity, although visibilities as high as 0.98 are obtained. As the SLED power is progressively reduced, the coherent heterodyne beat signals are increasingly dominated by uncorrelated noise contributions, leading to a gradual decrease in the maximum correlation. As illustrated in the inset of Fig.~\ref{fig:corr_100us}, for an acquisition time of \SI{100}{\micro\second}, the correlation peak corresponding to a SLED power of \SI{10}{\nano\watt} is already barely distinguishable from the noise floor (at $2\times10^{-3}$).

\subsection{Detection limit}\label{sec:detection_limit}

\begin{figure}[ht!]
\centering
\includegraphics[width=\hsize]{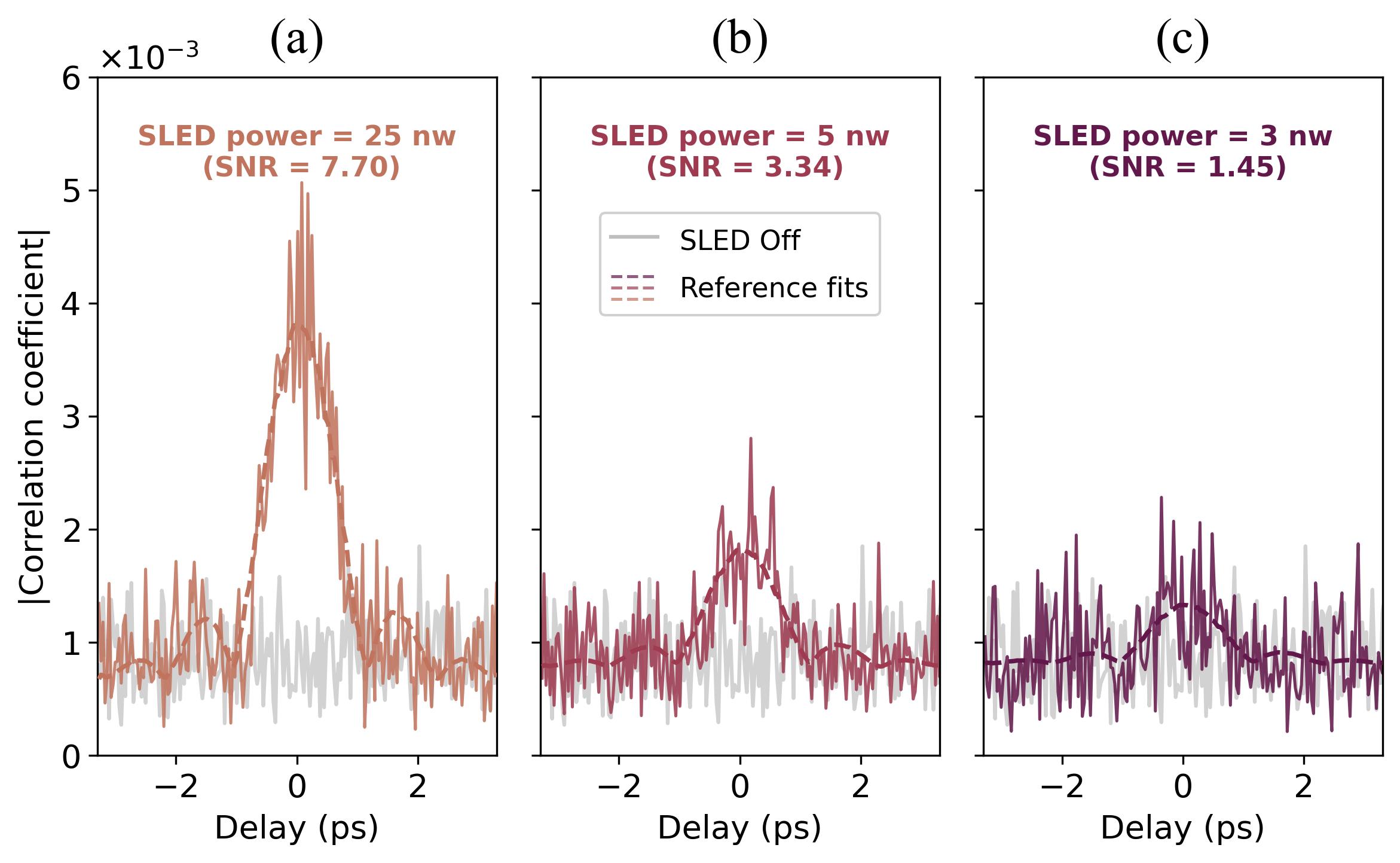}
\caption{Temporal correlation functions for nW SLED powers at oscilloscope’s maximum integration time of \SI{400}{\micro\second} sampled at \SI{100}{GS/s}.}
\label{fig:corr_400us}
\end{figure}

Since the contribution of uncorrelated noise averages down as the inverse square root of the acquisition time, the signal-to-noise ratio of the measured correlation peak improves proportionally to $\sqrt{T_{\mathrm{acq}}}$. To verify this scaling, the acquisition time is increased to \SI{400}{\micro\second}. Fig.~\ref{fig:corr_400us}a shows the correlation coefficient at low SLED power. As expected, the noise floor is reduced by a factor of two, leading to a twofold improvement in the measurement signal-to-noise ratio (SNR).
The SNR was estimated by fitting the measured correlation trace with a reference correlation function, acquired at high SLED power, using the amplitude of the best-fit model as the signal level. The noise level was then estimated as the standard deviation of the residuals between the measured data and the fitted model. The SNR is then defined as the ratio of the fitted amplitude to the noise level. 

In Fig.~\ref{fig:corr_400us}(c), an SNR exceeding 1.4 is achieved for a SLED power of \SI{3}{\nano\watt} and an integration time of $T=\SI{400}{\micro\second}$. This corresponds to an SNR of $\simeq 2$ under the conventional interferometric definition based on the squared fringe visibility. The SLED power is measured before the 1:2 split, giving a total spectral flux density below \SI{375}{\pico\watt\per\nano\meter} at \SI{1560}{\nano\meter}, or below \SI{188}{\pico\watt\per\nano\meter} per detection arm (i.e., per telescope) after the split. This per-telescope flux density can be  compared  with the expected flux density of a bright infrared source such as \emph{$\alpha$ Orionis}, which is approximately \SI{40}{\pico\watt\per\nano\meter} at \SI{1560}{\nano\meter} for a 1-m telescope, according to the 2MASS database.

\subsection{Shot noise limit}
\label{sec:shot_noise}
 
\begin{figure}[ht!]
\centering
\includegraphics[width=\hsize]{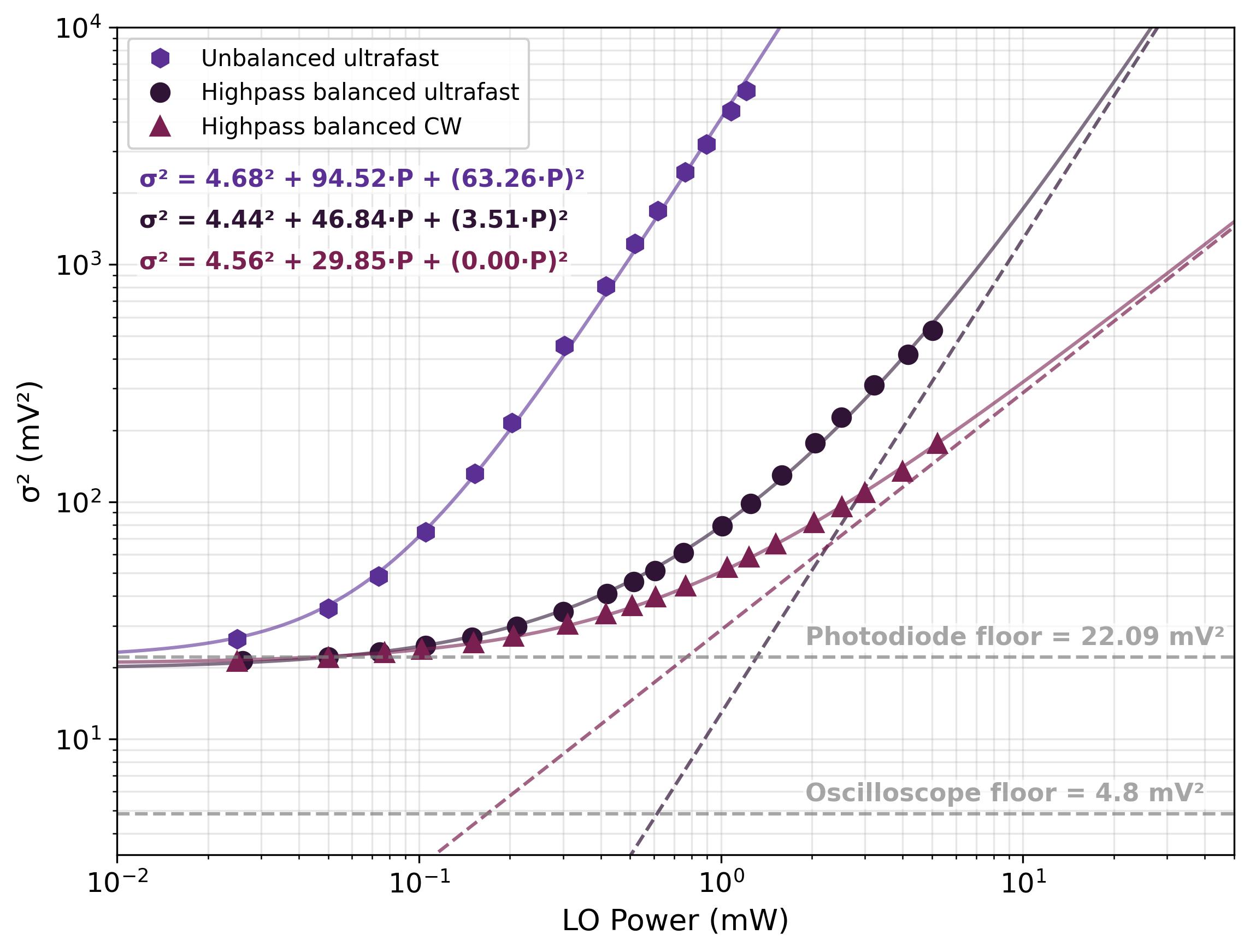}
\caption{Evolution of the voltage variance ($\sigma^2$) versus optical LO power for a shot-noise-limited CW laser and the ultrafast swept-source in different balancing configurations. Dashed lines indicate the expected scaling for a shot-noise-limited regime ($\sigma^2 \propto P$, slope 1) and a RIN-dominated regime ($\sigma^2 \propto P^2$, slope 2). The solid curves show weighted least-squares fits of Eq.~\ref{eq:noise_model} to the data.}
\label{fig:RMS2}
\end{figure}

Balanced heterodyne detection is expected to suppresses common-mode intensity fluctuations, and, in particular the relative intensity fluctuation/noise of the LO and to operate close to the shot-noise limit. However, the finite common mode rejection ratio (CMRR, not provided for the BPR-22) and/or imperfractions in the 50/50 couplers might prevent reaching this regime. In serrodyne detection, the LO exhibits strong classical intensity fluctuations because of the time-frequency mapping, and efficiency balanced detection and/or filtering of modulations at the repetition rate (here 80\,MHz) is essential. To assess whether the shot-noise regime is reached for LO power of 3\,mW, which is the maximum usable LO power, Fig.~\ref{fig:RMS2} presents the mean-square voltage fluctuations of one of the \SI{22}{\giga\hertz} amplified balanced photoreceivers as a function of the incident LO optical power (ie without SLED). For the sake of comparison, this measurement was reproduced: (a) with a large skew (\SI{20}{\pico\second}) between the two optical inputs of the balanced photodector, (b) without the high-pass filter and (c) by replacing the mode-locked laser by a low-noise single-frequency tunable laser (Tunics).

The experimental data were then fitted using a three-component noise model accounting for the distinct physical origins of the noise contributions:
\begin{equation}
    \sigma(P)^2 = \sigma_{\text{floor}}^2 + s_{\text{shot}} \cdot P + \left(s_{\text{class}} \cdot P\right)^2
    \label{eq:noise_model}
\end{equation}
where $\sigma_{\text{floor}}$ is a power-independent electronic noise floor set by the photodetectors and acquisition system, $s_{\text{shot}}$ is the shot-noise coefficient and $s_{\text{class}}$ is the classical/technical noise coefficient. The low-noise continuous laser increases as $P$ which confirms that its RIN is limited by shot-noise. With skew or without the high-pass RF filter, serrodyne signals increase directly as $P^2$ and the RIN is clearly not limited by shot-noise but by fast technical fluctuations. With proper balance detection (ie without skew) and with the high-pass RF filter, the RIN of the serrodyne signals is twice above the expected shot-noise limit, which explains the slightly higher noise floor in Fig.~\ref{fig:corr_100us} and Fig.~\ref{fig:corr_400us}.

\section{Discussion}\label{sec:discussion}

The experimental results demonstrate that serrodyne detection with a broadband mode-locked laser behaves essentially as well as heterodyne detection with a single-frequency laser with a small penalty (factor of 2) on the SNR linked, most probably, to the limited CMRR of our balanced photoreceivers. Although not implemented here, actively shaping the spectral profile of the LO should help reduce its technical relative intensity noise (RIN) to the shot-noise limit.
The minimal detectable SLED flux density would also benefit from splicing the numerous fiber interconnections, to reduce the transmission losses as well as 
back-reflections at the photoreceiver facets and fiber end-faces increasing the RIN of the LO.

The present work establishes the proof of principle of the proposed detection scheme. Building on these results, future work will focus on implementing wavelength multiplexing and parallel detection to fully exploit the potential of the technique in terms of acquisition speed and throughput. The straightforward parallelization afforded by the serrodyne detection scheme described in Section~\ref{Serrodyne principle} should enable the use of 16 CWDM spectral channels. Combined with an integration time of \SI{40}{\milli\second}, this approach is expected to achieve a sensitivity below \SI{10}{\pico\watt\per\nano\meter} in the H band could be achieved , even with uncooled photoreceivers.

Another key feature of serrodyne detection is its intrinsic compatibility with spectrally resolved observations. By scanning the group delay of the LO in one arm, spectra can be recovered in a manner analogous to Fourier-transform infrared spectroscopy (FTIR), with a spectral resolution ultimately limited by the instantaneous bandwidth of the serrodyne LO. This linewidth scales as $1/\sqrt{\phi_2}$ and, for the selected optical bandwidth of 1\,THz and a pulse duration of 12.5\,ns, is approximately 22\,GHz. This value matches the photoreceiver bandwidth and corresponds to a spectral resolving power of approximately 8,600.

\section{Conclusion}\label{sec:conclusion}
We have presented a new approach to optical heterodyne interferometry based on ultrafast lasers and fiber-optic telecom components, enabling scalable spectral multiplexing to circumvent bandwidth limitations. Although demonstrated here with a single spectral channel in the H band, the proposed architecture naturally extends to multi-channel operation and can be further expanded toward the K band by exploiting the broad bandwidth of mode-locked erbium-doped fiber lasers or subsequent spectral broadening in nonlinear fibers. The sensitivity achieved in this work, together with the availability of broadband telecommunication components, may provide a technological pathway toward large-scale spectrally multiplexed heterodyne interferometers with substantially improved sensitivity for the observation of protoplanetary disks and evolved stellar environments.

\bibliographystyle{aa}
\bibliography{heterodyne}


\end{document}